\documentclass[apjl]{emulateapj}

\usepackage{rotating}
\usepackage{graphicx}
\usepackage{epstopdf}

\newcommand{\msun}{~M$_\odot$}

\begin{document}

\title{The Response of Giant Stars To Dynamical-Timescale Mass Loss}
\author{Jean-Claude Passy$^{1,2}$, Falk Herwig$^1$, Bill Paxton$^3$}

\altaffiltext{1}{Department of Physics and Astronomy, University of Victoria, Victoria, BC, Canada}
\altaffiltext{2}{Department of Astrophysics, American Museum of Natural History, New York, NY, USA}
\altaffiltext{3}{Kavli Institute for Theoretical Physics, UC Santa Barbara, CA, USA}

\begin{abstract}

We study the response of giant stars to mass loss. One-dimensional simulations of red and asymptotic giant branch stars with mass loss rates from $10^{-3}$ up to a few \msun/yr show in no case any significant radius increase. The largest radius increase of 0.2\% was found in the case with the lowest mass loss rate. For dynamical-timescale mass loss rates, that may be encountered during a common envelope phase, the evolution is not adiabatic. The superadiabatic outer layer of the giant's envelope has a local thermal timescale comparable to the dynamical timescale. Therefore, this layer has enough time to readjust thermally. Moreover, the giant star is driven out of hydrostatic equilibrium and evolves dynamically. In these cases no increase of the stellar radius with respect to its initial value is found. If the mass loss rate is high enough, the superadiabaticity of the outer layer is lost progressively and a radiative zone forms due to a combination of thermal and dynamical readjustment. Conditions for unstable mass transfer based on adiabatic mass loss models that predict a significant radius increase, may need to be re-evaluated.

\end{abstract}

\keywords{binaries: close ---
		 binaries: general ---
		 methods: numerical ---
		 stars: evolution ---
		 stars: general ---
		 stars: mass loss}
		 
\section{Introduction}
\label{sec:intro}

Understanding how stars respond when they lose mass is a key ingredient on which binary evolution models depend. This response is particularly important in the context of interacting binaries with a donor filling its Roche lobe on a giant branch. In the oversimplifying case of conservative mass transfer, the orbital separation shrinks if the giant donor is more massive than the companion. If in the meantime the giant star expands or does not contract faster than the orbit shrinks, this positive feedback leads to an increase of the mass transfer rate. The stellar response to mass loss therefore dictates, along with how angular momentum is lost, whether or not a given system enters a common envelope phase \citep{Paczynski1976}. Consequently, it significantly affects the results of population synthesis studies \citep[see, e.g.,][]{PolitanoEtAl2010}.

Also, detailed 3D hydrodynamical simulations of the dynamical common envelope phase have shown that a giant's simulated envelope material is significantly lifted, but most of it does not reach escape velocity under the present modeling assumptions \citep[]{PassyEtAl2012,RickerTaam2012}. \cite{AlphaPaper2011} suggested that  an expansion of the giant as a result of mass loss \citep{HjellmingWebbink1987, GeEtAl2010} might contribute to the envelope ejection. Such an expansion of mass-losing giants was recently questioned in a \emph{Letter} by \cite{WoodsIvanova2011}. 

Therefore, we study the radius response of mass-losing giants again, with detailed microphysics, using the one-dimensional stellar evolution code MESA \citep[Module for Experiment in Stellar Astrophysics,][]{PaxtonEtAl2011}. Such a tool -- although it neglects three-dimensional effects -- allows to remove one or more simplifying assumptions adopted in some previous studies, which we mention in the next paragraphs.

For a star following a polytropic stratification of index $n$ with an adiabatic index $\gamma = 1+1/n$, the Lane-Emden equation leads to the {\it mass-radius relation} between a standard solution of radius $R_0$ and mass $M_0$, and a perturbed polytrope of radius $R$ and mass $M$:

\begin{equation}
	\frac{R}{R_0} = \left( \frac{M}{M_0} \right)^{\frac{1-n}{3-n}}.
	\label{eq:mass-radius}
\end{equation}

\noindent Note that Equation~(\ref{eq:mass-radius}) is only valid for a fixed adiabat throughout the stellar interior. For the complete derivation, see, e.g., \cite{HjellmingWebbink1987} or \cite{Carroll}. For an ideal gas equation of state, the specific entropy follows a simple expression:

\begin{equation}
	s(m) = s_0 + (1 + 1/n - \gamma)c_v \ln(\rho)
	\label{eq:entropy}
\end{equation}

\noindent where $s_0$ is a constant and $c_v$ is the specific heat at constant volume. Perfect monoatomic gases have $\gamma = 5/3$ so a convective region ($ds/dm = 0$) can be modeled with a polytrope of index $n = 3/2$. Using this value in Equation~(\ref{eq:mass-radius}) leads to $R/R_0 = (M/M_0)^{-1/3}$ and the conclusion that fully convective stars expand when they lose mass.

Later on, \cite{HjellmingWebbink1987} investigated the stability of polytropes, condensed polytropes (a polytropic envelope with a core modeled by a point mass) and composite polytropes (an envelope and a core with different polytropic indices) for convective ($\gamma = 1+1/n$) and radiative ($\gamma > 1+1/n$) regions. They evolved their models in the adiabatic regime, which means that they assumed hydrostatic equilibrium and an adiabatic evolution such that the entropy profile remains constant in Lagrangian coordinates. For the condensed polytropes, they showed (their Equation~40) that the {\it
adiabatic radius-mass exponent}, $\xi_{\rm ad}$, asymptotically approaches

\begin{equation}
	\xi_{\rm ad} \equiv \left(\frac{d \ln R }{d \ln M }\right)_{\rm ad} = \frac{1}{3-n}\left(1-n+\frac{m_c}{1-m_c}\right)
	\label{eq:xi}
\end{equation}

\noindent where $m_c$ is the ratio between the core mass and the total mass of the star. Equation~(\ref{eq:xi}) describes the behavior of $\xi_{\rm ad}$ in the limit $m_c \rightarrow 1$ but one also recovers the appropriate $\xi_{\rm ad}$ for a complete polytrope for any value of $n$ ($m_c=0$, Equation~\ref{eq:mass-radius}). However, the response of stars to very high mass loss rates may not be hydrostatic. Moreover, Equation~(\ref{eq:xi}) is only valid for condensed polytropes which are models that neglect radiation pressure and do not reproduce the superadiabatic regime that is encountered in the outer layers of giants.

\cite{GeEtAl2010} also studied the response of mass-losing stars in the adiabatic limit but used a detailed equation of state instead of a polytropic stratification. The star was assumed to stay in hydrostatic equilibrium, and its response to mass loss was assumed to be fully adiabatic. The entropy and composition profiles were fixed and the local value of these profiles during the evolution was obtained by interpolation from the initial model. For their 1\msun \ giant star model, the stellar radius increased by 30\%. The study concluded that instability in the mass transfer occurs rapidly for donors with a convective envelope, if at all, while donors with a radiative envelope may encounter a delayed dynamical instability. \cite{DeloyeTaam2010} used a similar approach to study the common envelope outcomes of 10\msun \ donors. They found that the mass of the remnant can vary by 20\% depending on when the common envelope phase happens and on the initial mass ratio.

All the models above yield a paradigm in which giants expand as a result of mass loss, such that mass loss that starts in semi-detached binaries with a giant donor tends to be unstable. However, for typical mass loss rates encountered at the onset of a common envelope interaction ($\dot{M} \la 1$\msun/yr, \citealt{PassyEtAl2012}), the donor does not stay in hydrostatic equilibrium. Moreover, an adiabatic evolution assumes that mass loss happens on a timescale shorter than the thermal timescale of the mass-losing star throughout its interior. 

Recently, \cite{WoodsIvanova2011} showed that the evolution could be locally non-adiabatic, since the outer superadiabatic layer of giant stars has a thermal timescale so short that it might readjust and reconstruct faster than it is stripped away. They present the evolutionary sequence of a 5\msun \ giant star for various mass loss rates, and show that the star grows mildly in radius during its evolution (their Figure~3). They also calculate the critical mass ratio for stable mass transfer for different donors. These values are only indicative but show that the response of the mass-losing star evolves with mass loss and cannot be parametrized using only the binary and stellar parameters.

In this paper we present models of mass-losing stars by removing some of the assumptions made in previous investigations. We obtain these models using the stellar evolution code MESA. The numerical method is described in Section~\ref{sec:numerical}. We present the simulations in Section~\ref{sec:simulations} and verify our method with low-mass zero-age main sequence (ZAMS) models in Section~\ref{sec:lowmass}. We then study in detail the dynamical response of red giant branch (RGB) and asymptotic giant branch (AGB) stars, and describe the physical processes involved, in Section~\ref{sec:giants}. A summary and conclusions are provided in Section~\ref{sec:summary}.

\section{Numerical method}
\label{sec:numerical}

MESA is a parallel one-dimensional stellar evolution code that uses adaptive mesh refinement and adaptive time stepping. In this section, we outline the basic features of this code. More details can be found in \cite{PaxtonEtAl2011}.

In hydrodynamic mode, the full set of differential equations of stellar evolution is solved in the Lagrangian description:

\begin{equation}
	v = r \frac{d\ln{r}}{dt}
\end{equation}

\begin{equation}
	\frac{d\ln{r}}{dm} = \frac{1}{4 \pi r^3 \rho}
\end{equation}

\begin{equation}
	\frac{dv}{dt} = - 4 \pi r^2 \frac{d P}{dm} - \frac{G m}{r^2}
	\label{eq:momentum}
\end{equation}

\begin{equation}
	\frac{d\ln{T}}{dm} = \frac{d\ln{P}}{dm} \nabla
	\label{eq:e_transport}
\end{equation}

\begin{equation}
	\frac{d l}{dm} = \epsilon_{\rm nuc} - \epsilon_{\rm \nu} - c_PT \left[  (1-\nabla_{\rm ad} \chi_T)\frac{d \ln{T}}{dt} - \nabla_{\rm ad} \chi_\rho \frac{d \ln{\rho}}{dt}   \right]
	\label{eq:e_conserve}
\end{equation}

\noindent where the mass $m$ is the independent variable and $r$, $\rho$, $P$, $T$, $l$, $\nabla \equiv d\ln{T} / d\ln{P}$, $\epsilon_{\rm nuc}$, $\epsilon_{\rm \nu}$, $c_P$, $ \nabla_{\rm ad} \equiv \left( d\ln{T} / d\ln{P} \right)_{s}$, $s$, $v$ and $G$ are the radius, the density, the pressure, the temperature, the luminosity, the temperature gradient, the nuclear energy generation rate, the neutrino loss rate, the specific heat at constant pressure, the adiabatic gradient, the specific entropy, the velocity and the gravitational constant, respectively. In addition, $\chi_T \equiv  \left( d\ln{P} / d\ln{T} \right)_{\rho}$ and $\chi_\rho \equiv  \left( d\ln{P} / d\ln{\rho} \right)_{T}$. To close this set of equations, we obtained the equation of state from a set of tables computed with the {\it FreeEOS}\footnote{http://freeeos.sourceforge.net/} code, developed by Alan Irwin, in the EOS4 configuration.

In order to improve numerical stability we use some artificial viscosity following the treatment by \citet[][their Equation 3]{WeaverEtAl1978}. Aside from providing better stability for the code, artificial viscosity had no effect on the evolution based on comparison of sequences carried out with or without artificial viscosity.

Different options have been explored for modeling stellar mass loss. We first directly set the mass loss rate $\dot{M}$ to a constant value regardless of the evolutionary stage. In order to show the robustness of the results, we also study how the models respond to a variable mass loss rate, for instance a scaled up wind model \citep{Reimers1975}:

\begin{equation}
	\dot{M} =  \eta_{\rm R} \times 4 \times 10^{-13} \left(\frac{L}{L_\odot}\right) \left(\frac{R}{R_\odot}\right) \left(\frac{M_\odot}{M}\right)~~~[M_\odot/{\rm yr}]
	\label{eq:reimers}
\end{equation}

\noindent where $L$ is the luminosity and $\eta_{\rm R}$ is a dimensionless constant. We select values of $\eta_{\rm R}$  to model very high mass loss rates. Typical models for RGB winds use values of $\eta_{\rm R}$ around 0.5.

\section{The simulations}
\label{sec:simulations}

We perform 21 MESA simulations (Table~\ref{tab:runs}). We first investigate the behavior of low-mass ZAMS stars (models 1 to 4) in order to verify our method through a comparison with the results of \cite{GeEtAl2010} for such models. We then study the stellar response of a 0.89\msun \ RGB star (models 5 to 9) in hydrodynamic mode for various mass loss rates. Models 10 to 13 are equivalent to models 5 to 8 except that they are carried out in hydrostatic mode, such that dynamical effects are isolated. Additional sequences for a 0.74\msun \ AGB star (models 14 and 15) and for a 5\msun \ RGB star (models 16 to 18) allow comparison with the results of  \cite{WoodsIvanova2011}. We also verify that changing the atmosphere boundary conditions from the ``simple atmosphere'' default option to the ``Eddington grey'' option (model 19) does not modify the outcome of our simulations \citep[these options are described in][]{PaxtonEtAl2011}. In order to verify that the initial response of the star is captured accurately in our simulations, we finally examine extra models (models 20 and 21) which are similar to models 8 and 17, but with an initial timestep smaller by an order of magnitude. Time-stepping automatically readjusts and we find no difference between the corresponding models.

\begin{deluxetable*}{cccccccc}
\tabletypesize{\footnotesize}
\tablewidth{0pt} 
\tablecaption{The main parameters for the simulations: the model number, the main sequence mass of the star ($M_{\rm MS}$), the stellar ($M_0$) and core ($M_c$) masses, radius ($R_0$), and luminosity ($L_0$) at the start of the mass loss phase, the mass loss rate or Reimers parameter ($\dot{M}$), and whether or not the model is carried out in hydrodynamic mode.}
\tablehead{
	   \colhead{Model}                                    &
           \colhead{$M_{\rm MS}$/M$_\odot$}                             &
           \colhead{$M_0$/M$_\odot$}                   &
           \colhead{$M_c$/M$_\odot$}                                       &
           \colhead{$R_0$/R$_\odot$}                                     &
           \colhead{$L_0$/L$_\odot$}                   &
           \colhead{$\dot{M}$}                                    &
           \colhead{Hydro}                                          }
\startdata
1 & 0.30 & 0.30 & - & 0.28 & $1.3\times10^{-2}$ & $10^{-2}$\msun/yr & No \\
	2 & 0.40 & 0.40 & - & 0.35 & $2.3\times10^{-2}$ & $10^{-2}$\msun/yr & No \\
	3 & 0.50 & 0.50 & - & 0.45 & $4.1\times10^{-2}$ & $10^{-2}$\msun/yr & No \\
	4 & 0.30 & 0.30 & - & 0.28 & $1.3\times10^{-2}$ & $0.1$\msun/yr & No \\
	\\
	5 & 1.00 & 0.89 & 0.41 & 102 & $1.20\times10^3$ & $10^{-3}$\msun/yr & Yes \\
	6 & 1.00 & 0.89 & 0.41 & 102 & $1.20\times10^3$ & $10^{-2}$\msun/yr & Yes \\
	7 & 1.00 & 0.89 & 0.41 & 102 & $1.20\times10^3$ & $0.1$\msun/yr & Yes \\
	8 & 1.00 & 0.89 & 0.41 & 102 & $1.20\times10^3$ & $1.0$\msun/yr & Yes \\
	9 & 1.00 & 0.89 & 0.41 & 102 & $1.20\times10^3$ & $\eta_{\rm R} = 10^6$  & Yes \\
	\\
	10 & 1.00 & 0.89 & 0.41 & 102 & $1.20\times10^3$ & $10^{-3}$\msun/yr & No \\
	11 & 1.00 & 0.89 & 0.41 & 102 & $1.20\times10^3$ & $10^{-2}$\msun/yr & No \\
	12 & 1.00 & 0.89 & 0.41 & 102 & $1.20\times10^3$ & $0.1$\msun/yr & No \\
	13 & 1.00 & 0.89 & 0.41 & 102 & $1.20\times10^3$ & $1.0$\msun/yr & No \\
	\\
	14 & 1.00 & 0.74 & 0.52 & 111 & $1.37\times10^3$ & $0.5$\msun/yr & Yes \\
	15 & 1.00 & 0.74 & 0.52 & 111 & $1.37\times10^3$ & $1.0$\msun/yr & Yes \\
	\\
	16 & 5.00 & 4.99 & 0.61 & 50 & $8.85\times10^2$ & $10^{-2}$\msun/yr & Yes \\
	17 & 5.00 & 4.99 & 0.61 & 50 & $8.85\times10^2$ & $1.0$\msun/yr & Yes \\
	18 & 5.00 & 4.99 & 0.61 & 50 & $8.85\times10^2$ & $\eta_{\rm R} = 2\times10^7$ & Yes \\
	\\
	19\tablenotemark{a} & 1.00 & 0.89 & 0.41 & 102 & $1.20\times10^3$ & $0.1$\msun/yr & Yes \\
	20\tablenotemark{b} & 1.00 & 0.89 & 0.41 & 102 & $1.20\times10^3$ & $1.0$\msun/yr & Yes \\
	21\tablenotemark{c} & 5.00 & 4.99 & 0.61 & 50 & $8.85\times10^2$ & $1.0$\msun/yr & Yes \\
\enddata
\label{tab:runs}
\tablenotetext{a}{Similar to model 7 but with different boundary conditions.}
\tablenotetext{b}{Similar to model 8 but with an initial timestep ten times smaller.}
\tablenotetext{c}{Similar to model 17 but with an initial timestep ten times smaller.}

\end{deluxetable*}

\section{Low-mass zero age main sequence stars}
\label{sec:lowmass}

\begin{figure}[h!]
	\begin{center}
		\includegraphics[scale=0.3]{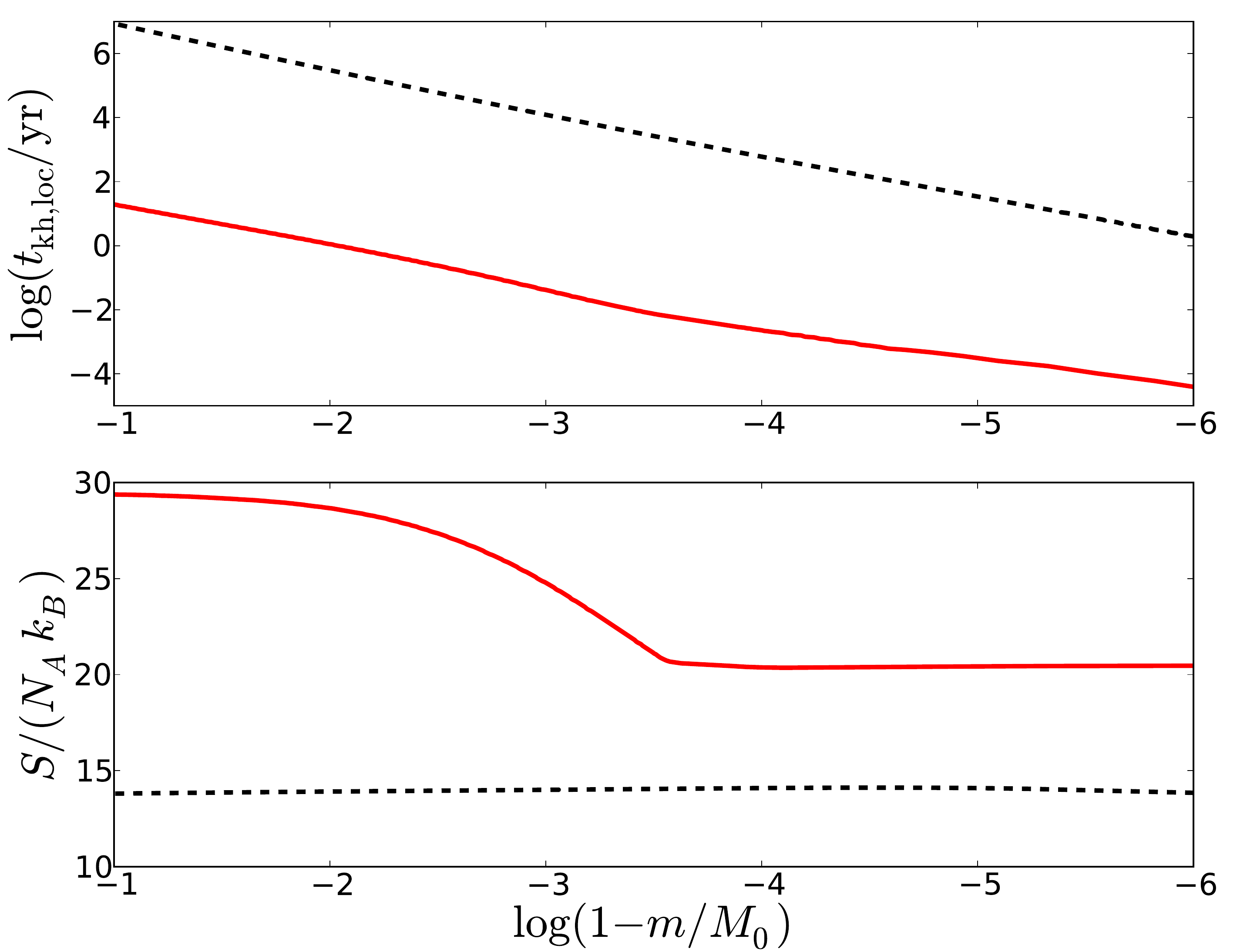}
	\caption{The local thermal timescale (top) and specific entropy (bottom) in the outermost 10\% of the stellar mass for the $M_0=0.3$\msun \ ZAMS star (dashed black, models 1 and 4) and the $M_0=0.89$\msun \ RGB star (solid red, models 5 to 8 and 14).
	\label{fig:ZAMSthloc}
	}
	\end{center}
\end{figure}

Low-mass ZAMS stars provide a simple case to compare our method --- solving the full set of stellar equations --- with the method used in \cite{GeEtAl2010} --- assuming an adiabatic evolution (c.f. Section~\ref{sec:intro}). Indeed, the global thermal timescale of a 0.3\msun \ ZAMS star is

\begin{equation}
	t_{\rm KH} \equiv \frac{G M^2}{RL} \approx 7.7 \times 10^8~{\rm years},
\end{equation}

\noindent which is several orders of magnitude longer than the duration of the different sequences for the mass loss rates considered (Table~\ref{tab:runs}). Consequently, the star does not have enough time to readjust thermally throughout its entire interior. Moreover, one can calculate the local thermal timescale of the outer part of the star \citep{WoodsIvanova2011}:

\begin{equation}
	t_{\rm KH,loc}(m) \equiv \int_m^M u(m^{'})/L(m^{'})~dm^{'}
	\label{eq:thloc}
\end{equation}

\noindent where $u$ is the internal energy per unit mass and $m$ is the mass coordinate. Equation~(\ref{eq:thloc}) implicitly assumes that one can parse the stellar interior at any mass coordinate into an outer and an inner zone, and that the outer zone can thermally readjust independently. This might not strictly be the case in a convective region where convective overturn is usually shorter than the thermal timescale of a given zone, thus stabilizes thermal perturbations on a timescale shorter than the thermal timescale of the zone. Figure~\ref{fig:ZAMSthloc} shows that even the outermost 0.01\% of the mass needs about 1000\,years to thermally readjust to the perturbations induced by mass loss. Again, this timescale is much longer than the duration of the simulations (25 years for the 0.3\msun \ ZAMS star with a mass loss rate of $10^{-2}$\msun/yr, model 1), and so the outer layers cannot thermally readjust. The evolution of the mass-losing star is thus adiabatic, both locally and globally. Among low-mass ZAMS stars, those with lower mass have deeper convective envelopes. The lowest-mass ZAMS stars ($M\la 0.3$\msun) are fully convective, and so should be well approximated by a polytropic expansion ($\xi_{\rm ad} = -1/3$, Equation~\ref{eq:xi}).

\begin{figure}[h!]
	\begin{center}
		\includegraphics[scale=0.34]{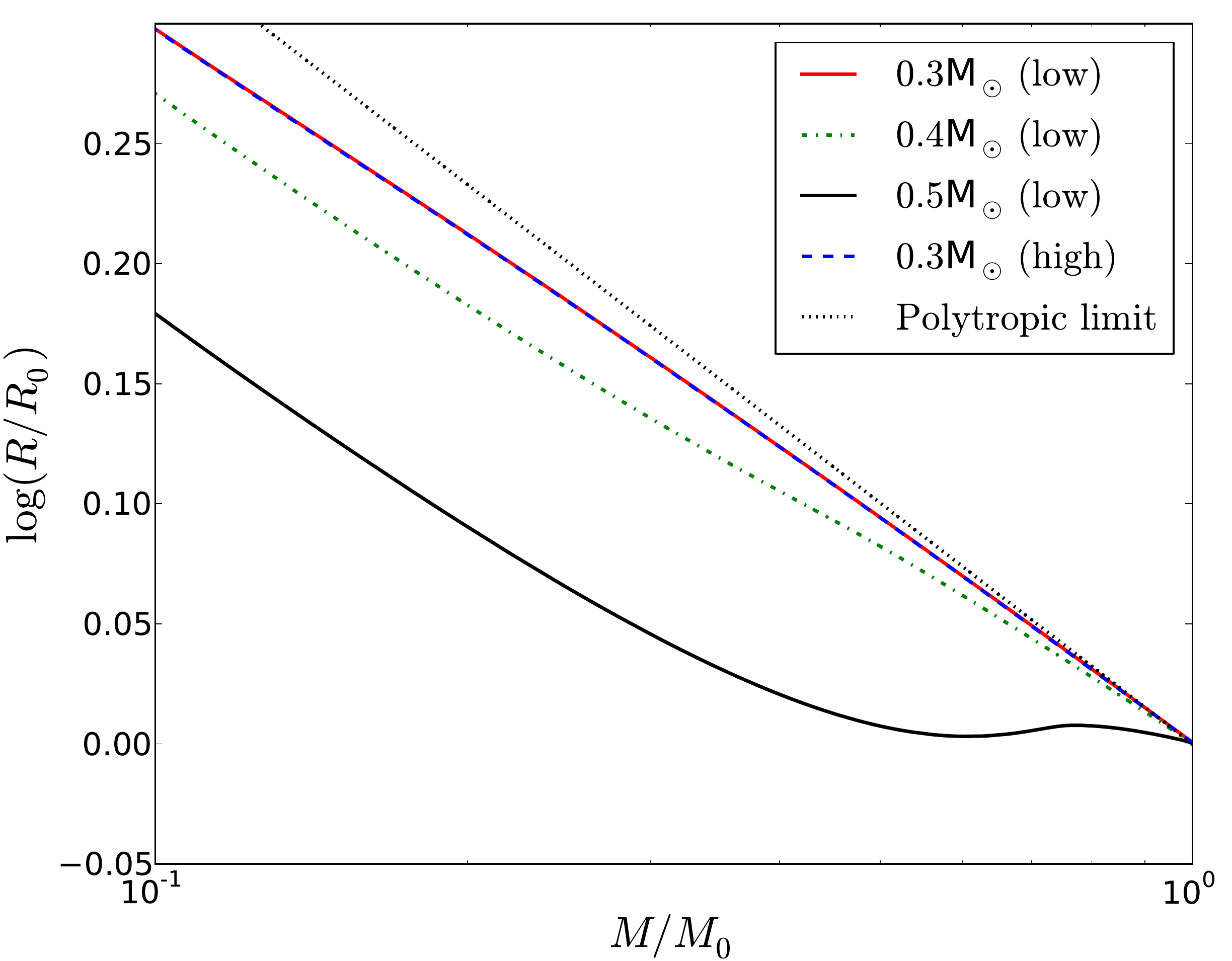}
	\caption{The evolution of the stellar radius as a function of stellar mass in terms of the initial radius ($R_0$) and mass ($M_0$) of the ZAMS models (1 to 4) with mass loss rates $\dot{M} = 10^{-2}$\msun/yr (low) and $\dot{M} = 0.1$\msun/yr (high). Also shown is the evolution in the polytropic limit (Equation~\ref{eq:mass-radius}).
	\label{fig:ZAMSEvolution}
	}
	\end{center}
\end{figure}

We carry out simulations where the mass-losing star is a ZAMS star with a mass ranging from 0.3 to 0.5\msun \ (models 1 to 4). The resolution is approximately 1000 zones. We reproduce the results from \citealt{GeEtAl2010} quite accurately (their Figure~3). For all models, the radius of the mass-losing star increases as mass is lost (Figure~\ref{fig:ZAMSEvolution}), in particular the 0.3\msun \ model, which is almost fully convective and therefore behaves most like a polytrope (Equation~\ref{eq:mass-radius}). One should emphasize that the small difference between the 0.3\msun \ model and the polytropic limit arise from deviations of the stellar structure from a complete polytrope. We try different mass loss rates of $10^{-2}$ and $10^{-1}$\msun/yr for the 0.3\msun \ star (models 1 and 4, respectively) and find no difference whatsoever. The evolution for both mass loss rates is much too rapid to allow any local thermal readjustment of the outer parts of the star (Figure~\ref{fig:ZAMSthloc}). In conclusion, the evolution of low-mass ZAMS stars is fully adiabatic. Thus, the approximation by \cite{GeEtAl2010} is appropriate for these low-mass main-sequence stars. 

\section{Giant stars}
\label{sec:giants}

The case of giant stars is somewhat more complicated. Giant stars have a convective envelope in which the entropy profile is flat. For intermediate-mass giant stars, the global thermal timescale of the star is still large in comparison with its dynamical timescale, and so the stellar interior cannot thermally readjust during most simulations. However, intermediate-mass giants also possess a cool, low-density outer layer in which convection is very inefficient. The local thermal timescale of this superadiabatic layer is very short in comparison with the global thermal timescale of the star: the outermost 1\% of the mass thermally readjusts in approximately one year (Figure~\ref{fig:ZAMSthloc}). Therefore, the layer might have enough time to thermally readjust locally, depending on how the local thermal timescale of the superadiabatic layer and the time needed to strip it away compare. Assuming that the layer has a mass $m_{\rm shell}$, its local thermal timescale is $t_{\rm KH,loc}(m_{\rm shell})$ and it takes $m_{\rm shell}/\dot{M}$ to remove it. This leads to a critical mass loss rate:

\begin{equation}
	\dot{M}_{\rm crit} \approx \frac{m_{\rm shell}}{t_{\rm KH,loc}(m_{\rm shell})}
	\label{eq:mdotcrit}
\end{equation}

\noindent which gives the threshold for the readjustment of the superadiabatic layer. If $\dot{M} \ll \dot{M}_{\rm crit}$, the superadiabaticity cannot be removed and the outer layer thermally readjusts. If $\dot{M} \ga \dot{M}_{\rm crit}$, the outer layer does not have time to readjust and superadiabaticity is lost progressively. The higher the mass loss rate, the sooner the superadiabatic layer disappears entirely, after which the star evolves adiabatically. A similar argument has been made by \cite{WoodsIvanova2011}, except that we consider here the time required to remove the superadiabatic layer rather than the time to strip away the entire star, as they did in their Equation~3. There is no unique definition of this outer layer, but for our 0.89\msun \ RGB model one can estimate from the entropy profile that $m_{\rm shell}$ is about $10^{-3}$\msun \ and $t_{\rm KH,loc}(m_{\rm shell}) \approx 0.04$~year, which leads to a critical value for the mass loss rate $\dot{M}_{\rm crit} \approx 2.5\times10^{-2}$\msun/yr.

\vspace{0.5cm}

\subsection{The canonical case of a 0.89\msun \ red giant branch star}

In this section, we first study the canonical case of a 0.89\msun \ RGB star, and compare our results to the adiabatic models from \cite{GeEtAl2010}. We carry out hydrodynamic simulations with constant mass loss rates ranging from $10^{-3}$ to 1\msun/yr (models 5 to 8) and one model with a varying mass loss rate (model 9). We also carry out their hydrostatic counterparts (models 10 to 13) in order to compare with previously published models, and to demonstrate the error a hydrostatic assumption causes. The resolution for all the one-dimensional models discussed here is approximately 2500 zones.

\begin{figure}[h!]
	\begin{center}
		\includegraphics[scale=0.39]{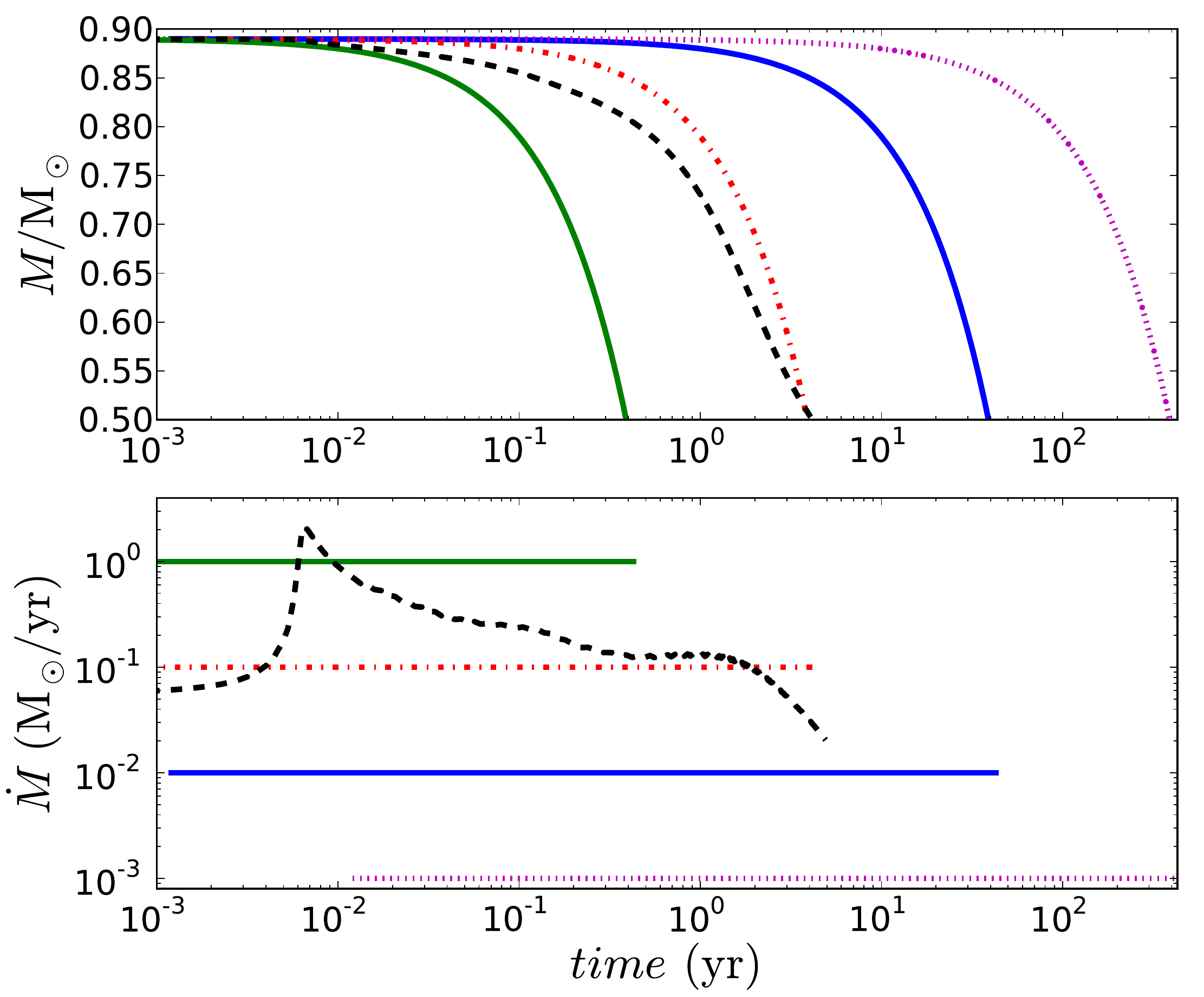}
	\caption{The evolution of the mass (top) and the mass loss rate (bottom) for the 0.89\msun \ RGB star with mass loss rates of $10^{-3}$\msun/yr (dotted magenta, model 5), $10^{-2}$\msun/yr (solid blue, model 6), $0.1$\msun/yr  (dash-dotted red, model 7), $1$\msun/yr (solid green, model 8), and variable (dashed black, model 9).
	\label{fig:1M_RGB_Mass}
	}
	\end{center}
\end{figure}

\begin{figure}[h!]
	\begin{center}
		\includegraphics[scale=0.37]{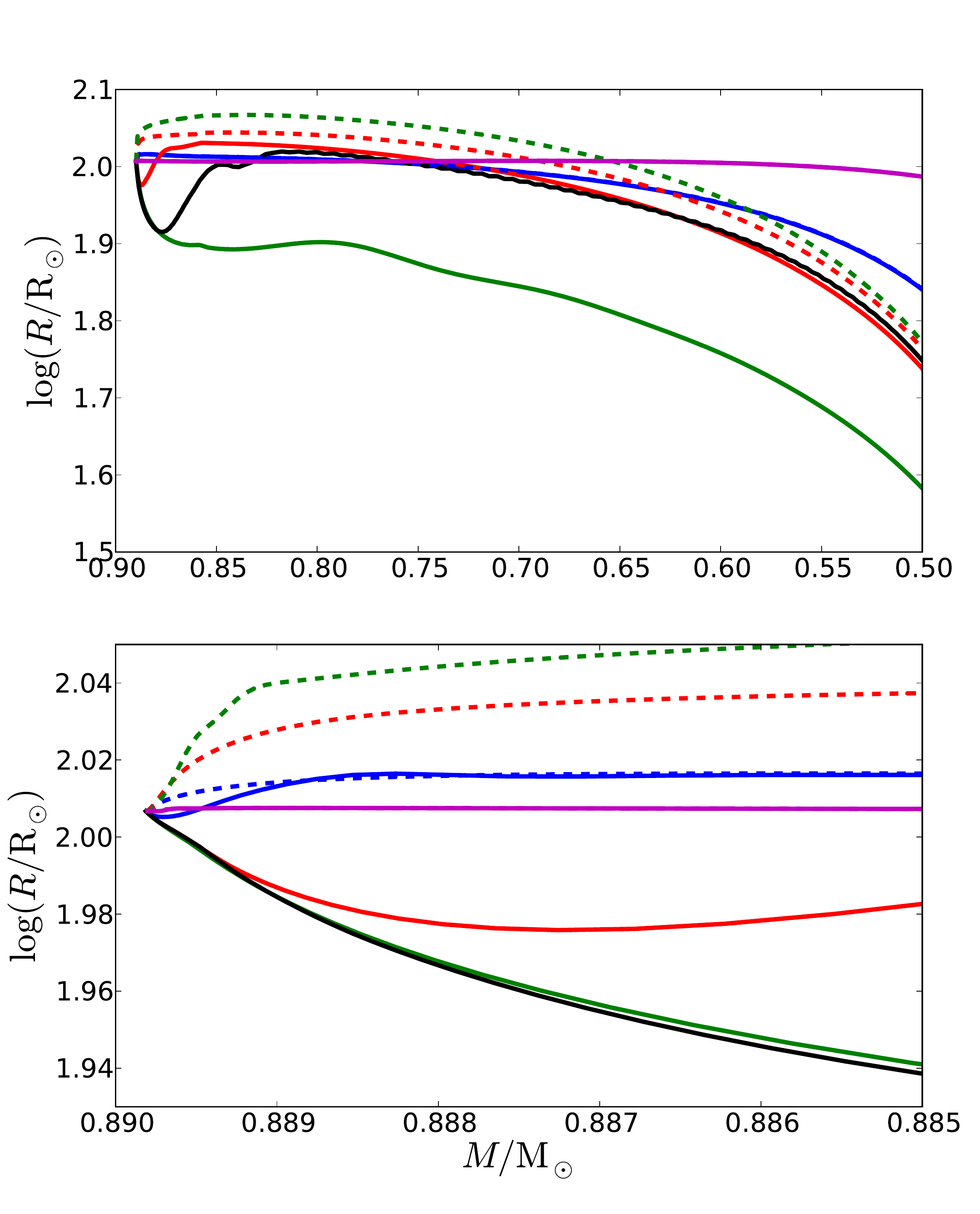}
	\caption{Top: evolution of the stellar radius as a function of stellar mass for the 0.89\msun \ RGB star with mass loss rates of $10^{-3}$\msun/yr (magenta), $10^{-2}$\msun/yr (blue), $0.1$\msun/yr  (red), $1$\msun/yr (green), and variable (black). Both hydrodynamic (solid) and hydrostatic sequences (dashed) are shown. Bottom: a close-up of the early evolution of the sequences shown above.
	\label{fig:1M_RGB_Radius}
	}
	\end{center}
\end{figure}

\begin{figure}[h!]
	\begin{center}
		\includegraphics[scale=0.38]{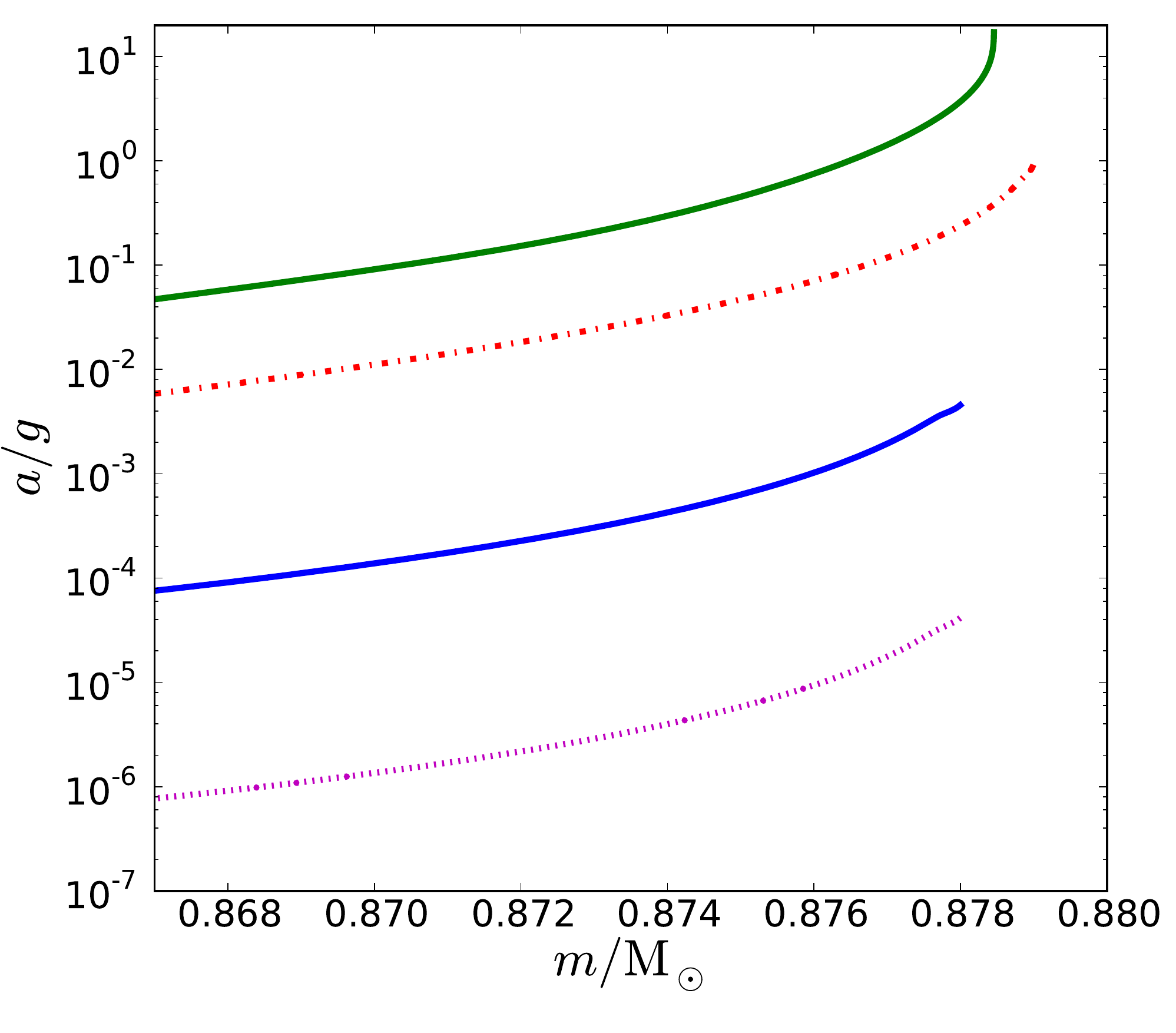}
	\caption{Ratio between the acceleration and the gravitational acceleration after the 0.89\msun \ RGB star star has lost about $10^{-2}$\msun. The mass loss rate is (from bottom to top) $10^{-3}$\msun/yr (dotted magenta, model 5), $10^{-2}$\msun/yr (solid blue, model 6), $0.1$\msun/yr  (dash-dotted red, model 7), and $1$\msun/yr (solid green, model 8).
	\label{fig:accel}
	}
	\end{center}
\end{figure}

We plot in Figure~\ref{fig:1M_RGB_Mass} the evolution of the stellar mass and the mass loss rate for models 5 to 9, while the evolution of the stellar radius for models 5 to 14 is shown in Figure~\ref{fig:1M_RGB_Radius}. Mass-losing giants barely expand, if at all. The difference between the hydrodynamic and the hydrostatic models for the lowest mass loss rate ($\dot{M} = 10^{-3}$\msun/yr, models 5 and 10) is hardly noticeable. This is due to the fact that for such a low mass loss rate, the star is barely driven out of hydrostatic equilibrium. A comparison between the acceleration ($a=dv/dt$) and the gravitational acceleration ($g=Gm/r^2$) profiles confirms that this model stays in hydrostatic equilibrium (Figure~\ref{fig:accel}). If one increases the mass loss rate to $10^{-2}$\msun/yr, some marginal differences arise in the very early phase of evolution between the hydrodynamic and the hydrostatic models. This is the threshold for which hydrodynamic effects can no longer be neglected, as the acceleration in the outer parts of the giant represents almost 1\% of the gravitational force at that location. These effects naturally increase as the mass loss rate increases.

Hydrodynamic models for which dynamical effects cannot be neglected (models 6 to 9) all contract in the early evolutionary phase. The higher the mass loss rate, the more the stellar radius decreases (Figure~\ref{fig:1M_RGB_Radius}). Later on, differences arise as stars with high mass loss rates keep contracting while stars with lower $\dot{M}$ first re-expand slightly and then contract again. For the model with $\dot{M} = 10^{-2}$\msun/yr (model 6), the radius of the star grows by less than 5\%. We  certainly do not see the $30$\% expansion found in \cite{GeEtAl2010} for a 1\msun \ giant star (their Figure~6). Model 8, for which the shrinkage is most dramatic, considers a typical mass loss rate that is encountered during a common envelope evolution. On the other hand, the equivalent hydrostatic ``test'' models (models 11 to 13) all show an expansion of the radius, with the higher mass loss rates leading to the largest increases. This behavior is similar to sequences by \citet[their Figure~3]{WoodsIvanova2011}. This comparison demonstrates that dynamical aspects play a critical role in the stellar response of our model. Some energy that would be transformed into internal energy or expansion work in the hydrostatic assumption can now go into kinetic energy.

\begin{figure}[h!]
	\begin{center}
		\includegraphics[scale=0.27]{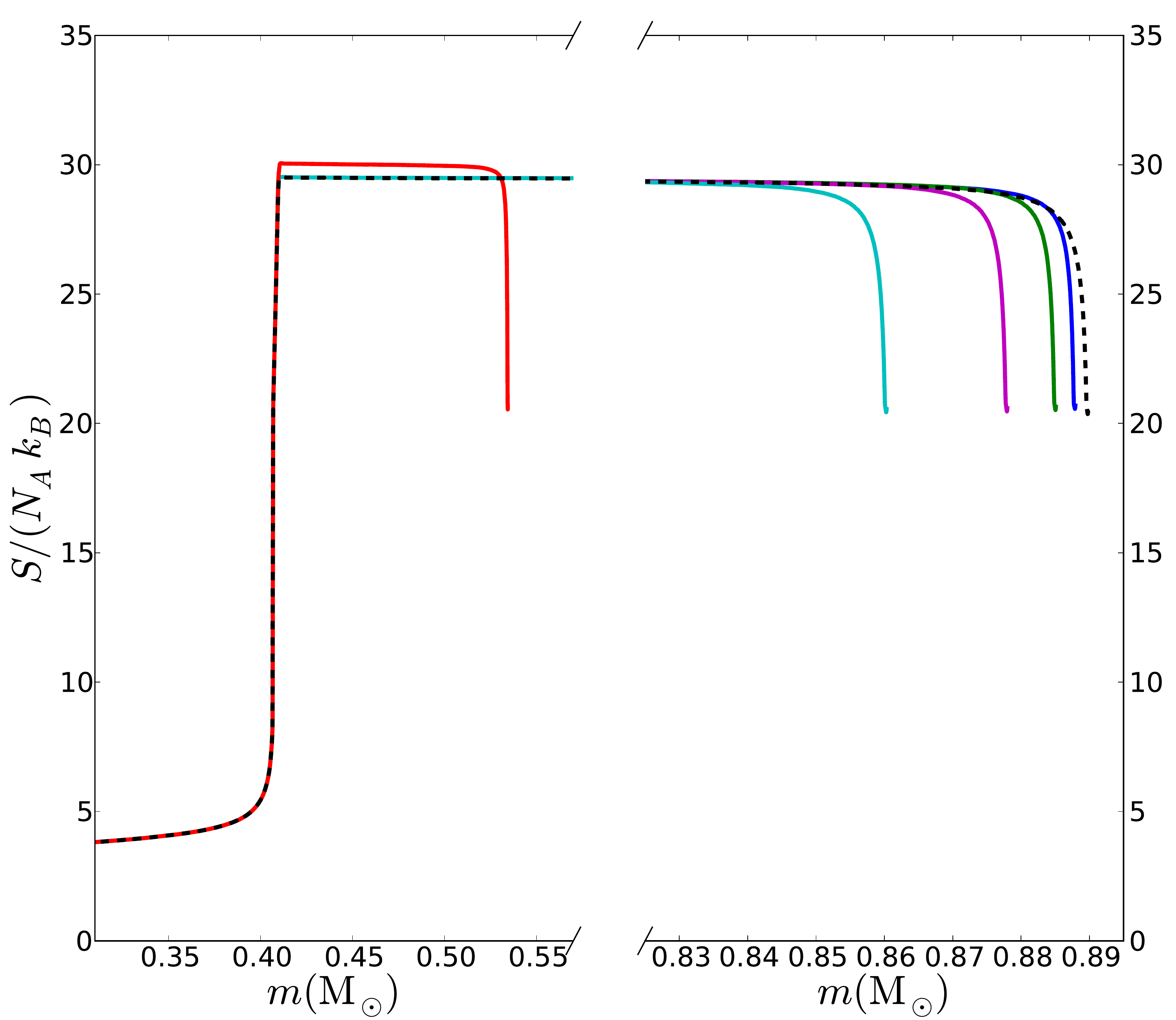}
		\includegraphics[scale=0.27]{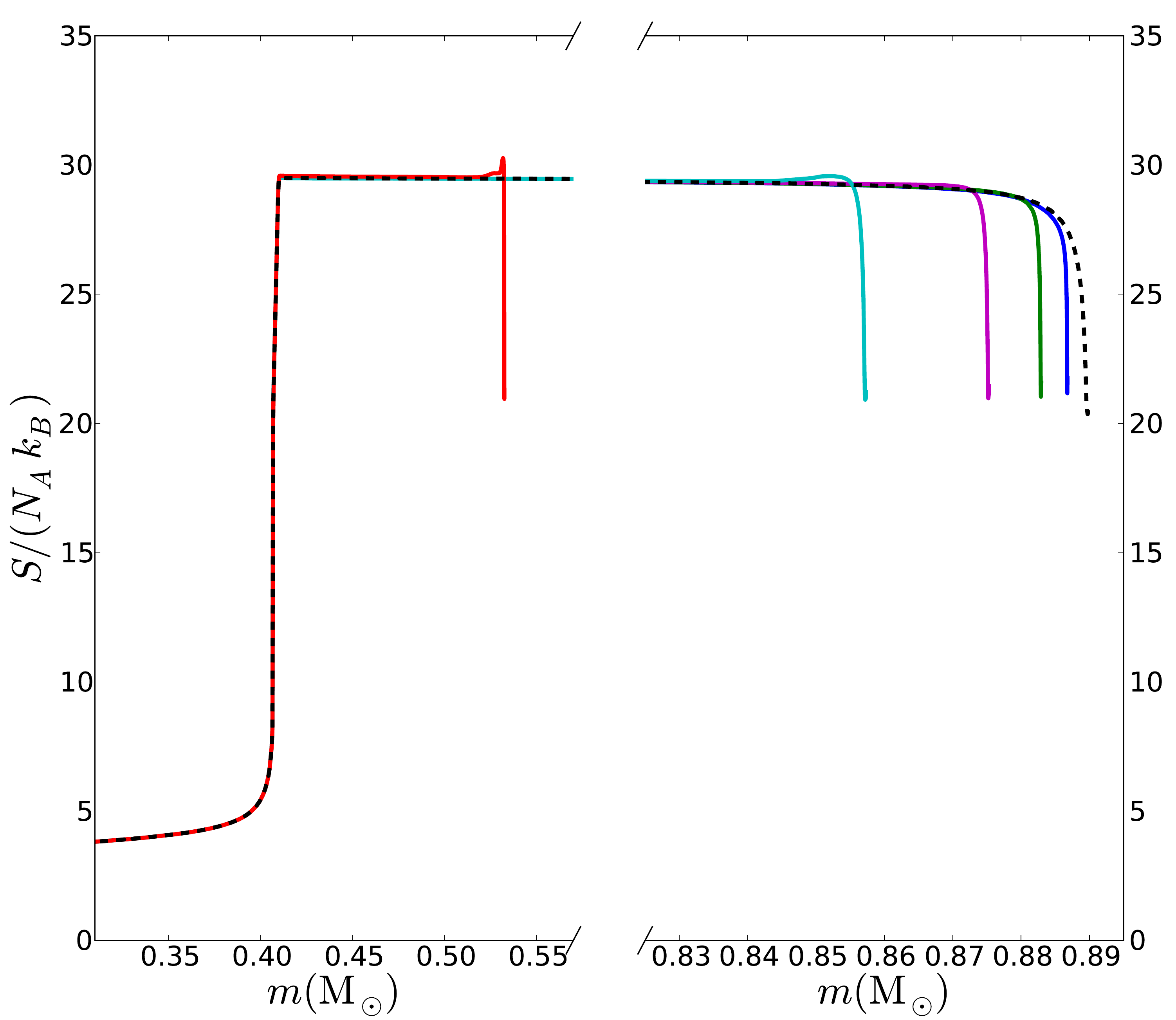}
		\includegraphics[scale=0.27]{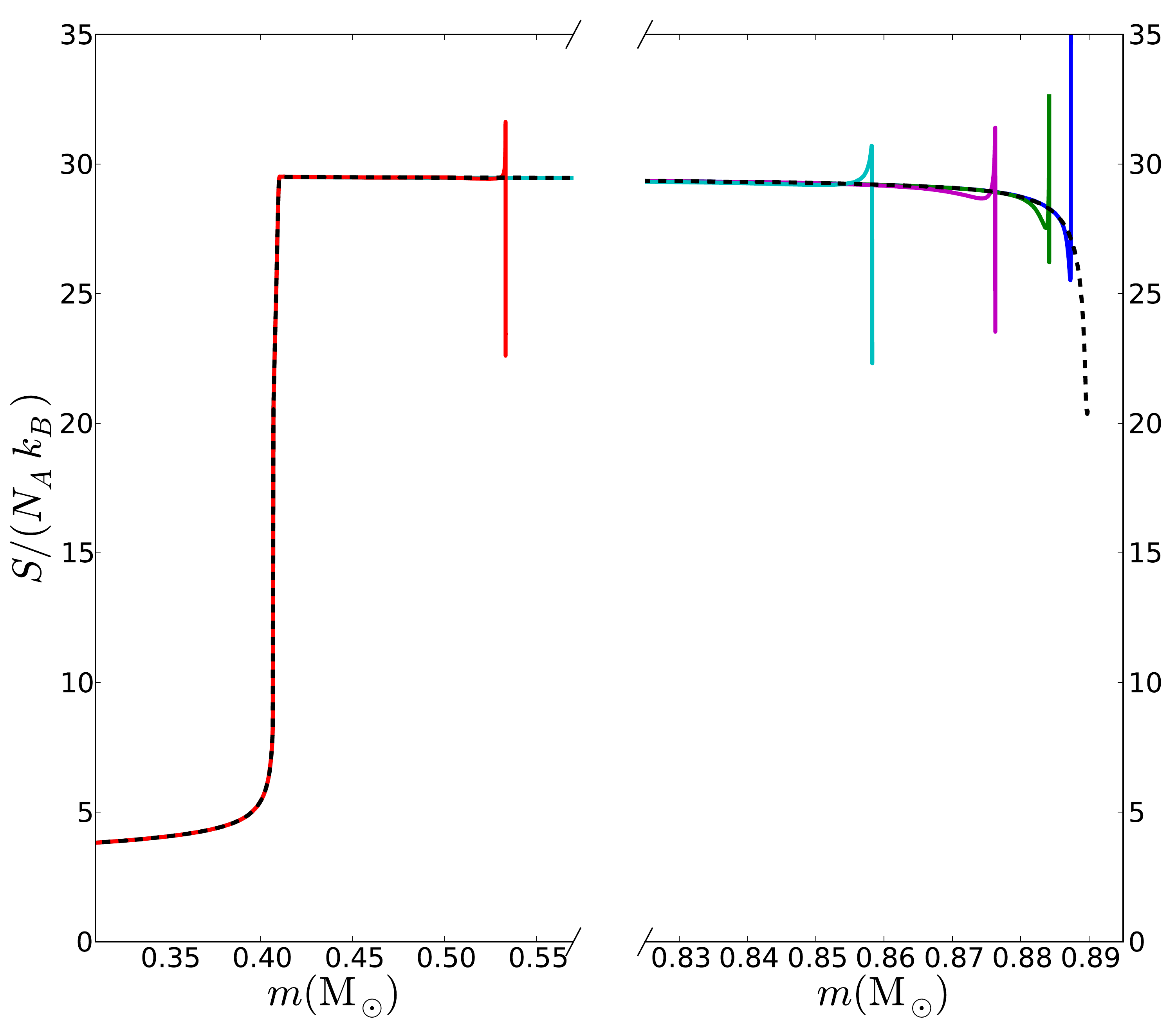}
	\caption{Entropy profiles at the onset of mass loss (dashed black), during the early evolutionary phase (solid colors, right panels) and after the 0.89\msun \ RGB star has lost about 40\% of its mass (red, left panels). The mass loss rate is $10^{-2}$\msun/yr (model 6, top), 0.1\msun/yr (model 7, middle) and 1\msun/yr (model 8, bottom).
	\label{fig:1M_RGB_Entropy}
	}
	\end{center}
\end{figure}

In order to understand the reasons for these different behaviors, we plot in Figure~\ref{fig:1M_RGB_Entropy} entropy profiles at different times for models 6, 7 and 8. The evolution of the entropy differs significantly between cases with different mass loss rates. For the lowest mass loss rate ($\dot{M} = 10^{-2}$\msun/yr, model 6), the entire interior has enough time to adjust thermally. After 40\% of the initial stellar mass has been lost, the star has still the entropy profile similar to the one of a giant star stratification. For the intermediate mass loss rate ($\dot{M} = 0.1$\msun/yr, model 7), very little of the outer layer loses its superadiabaticity in the early phase. Eventually, the interior of the star has not changed except in the outermost parts. Moreover, the superadiabatic layer is much less prominent than for a regular giant star stratification, and a radiative zone has developed just beneath it. In the highest mass loss rate case ($\dot{M} = 1$\msun/yr, model 8), one can see the superadiabaticity being removed very early in the evolution and the build-up of a radiative zone. Eventually, the surface is not superadiabatic anymore. The entropy profile of the interior layers has not changed and the mass for which the entropy drops is negligible. Only a small radiative zone remains on top of the convective zone (Figure~\ref{fig:1M_RGB_Entropy}, middle and bottom panels).

\begin{figure}[h!]
	\begin{center}
		\includegraphics[scale=0.43]{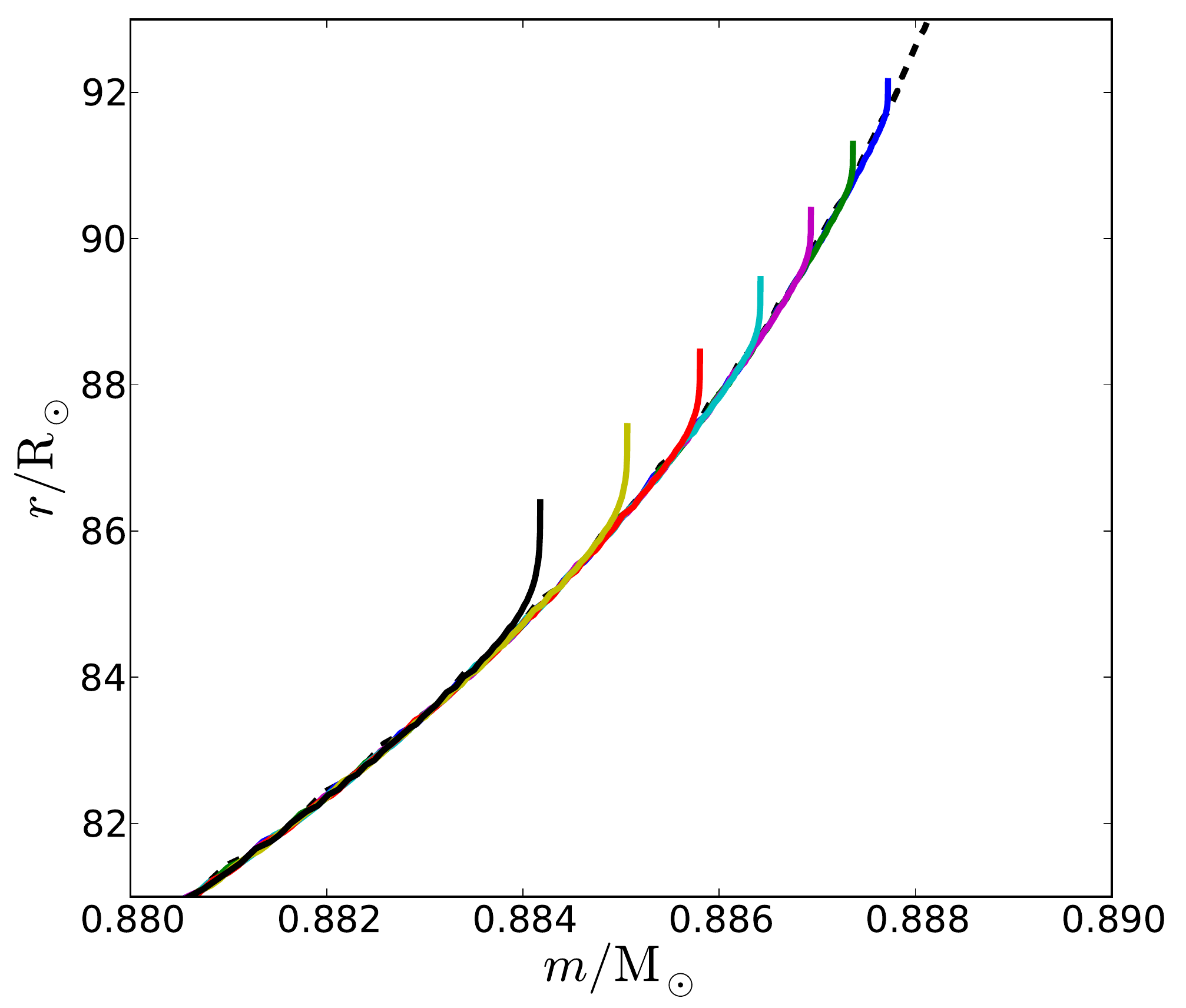}
		\includegraphics[scale=0.43]{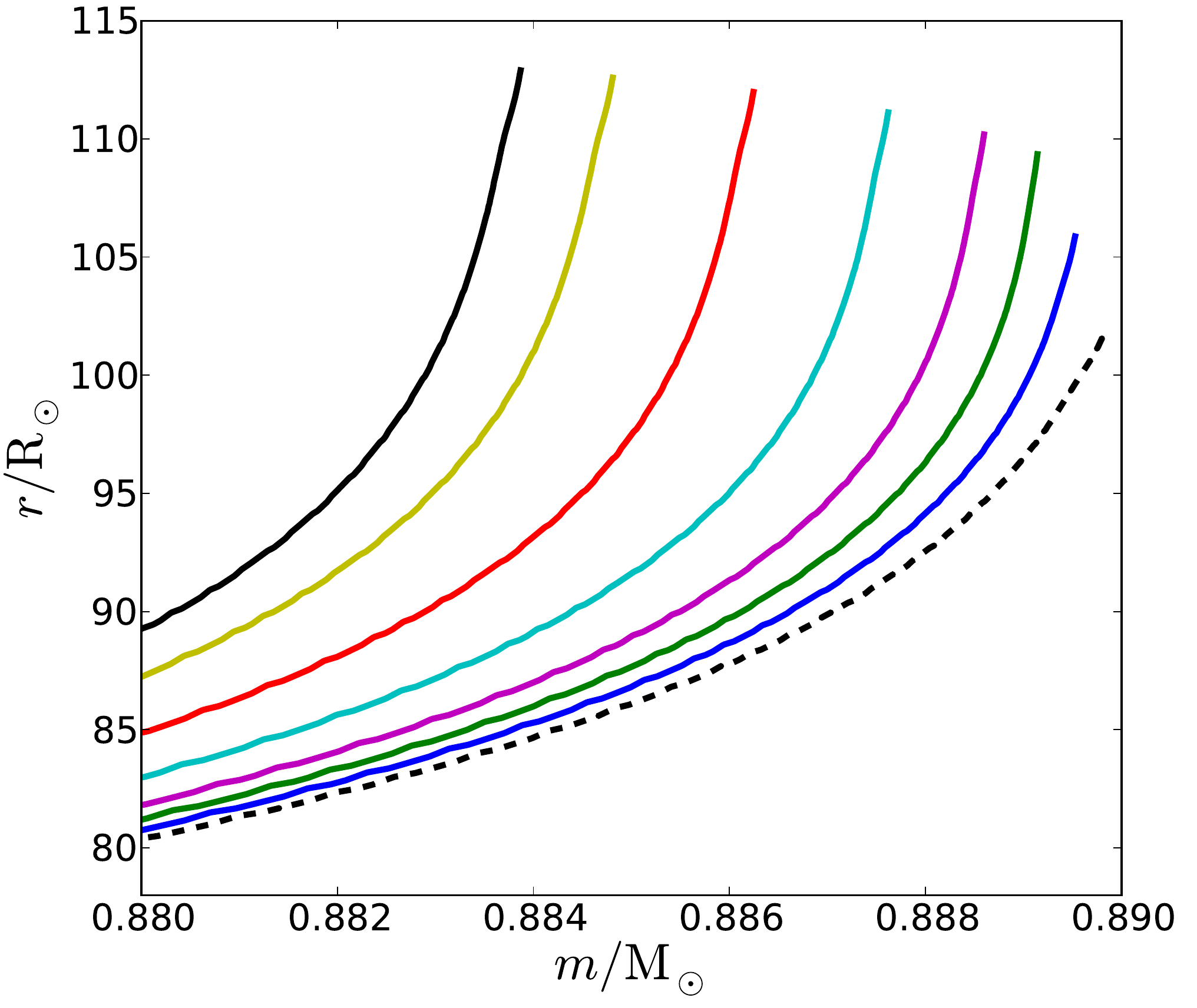}
	\caption{Radius profiles at the onset of mass loss (dashed black) and during the early evolutionary phase (solid colors) for the 0.89\msun \ RGB star and a mass loss rate of 1\msun/yr. Both hydrodynamic (top, model 8) and hydrostatic sequences (bottom, model 13) are shown.
	\label{fig:radius}
	}
	\end{center}
\end{figure}

The evolution of the radius profile as a function of mass for the hydrodynamic case shows that while some material close to the surface always moves out, the radius continuously decreases (Figure~\ref{fig:radius}). A certain kinetic energy is associated with this local outward motion which, in the case of hydrostatic models, goes into expansion work (potential energy). As a result the latter models increase their radius.

\begin{figure}[h!]
	\begin{center}
		\includegraphics[scale=0.34]{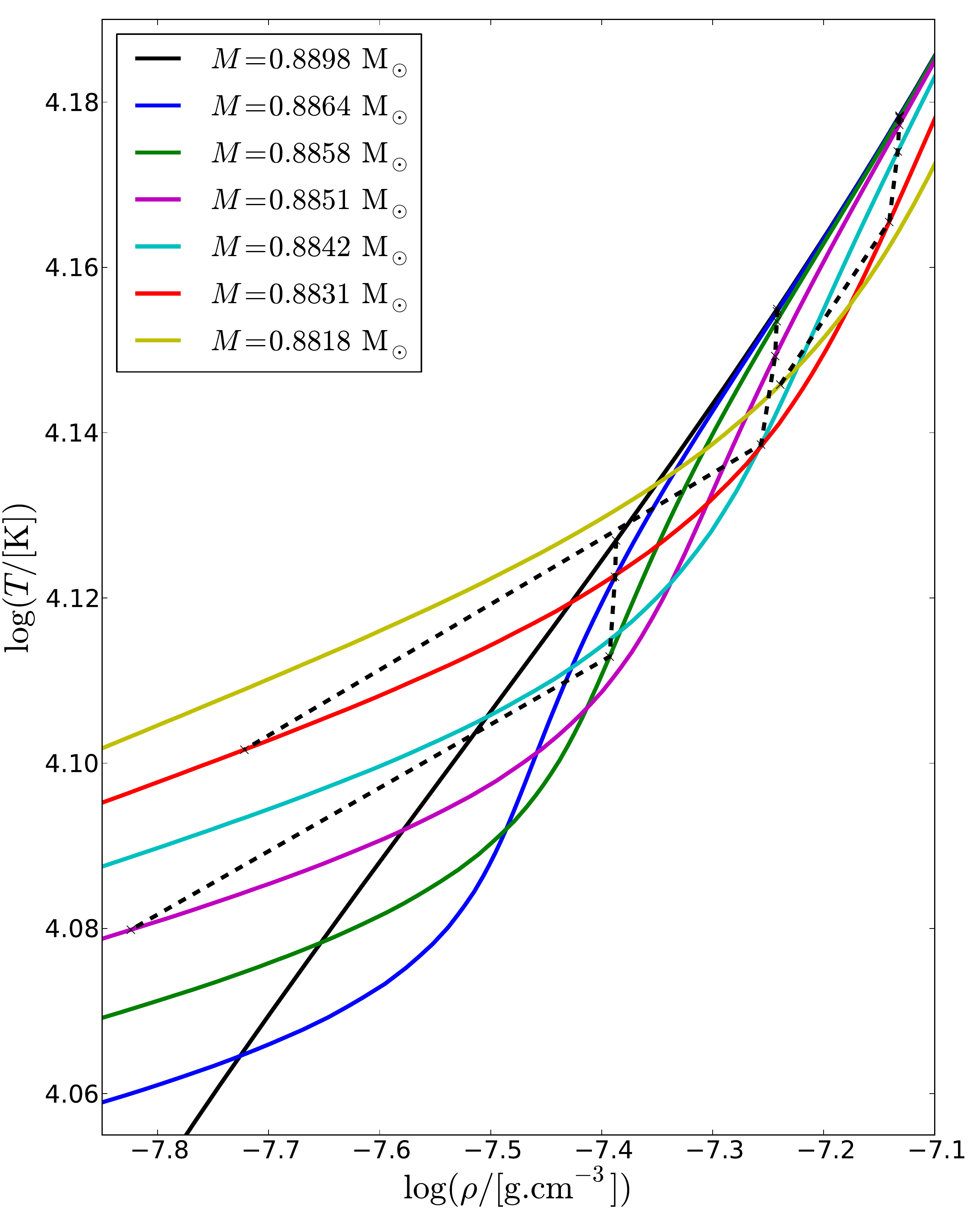}
	\caption{Profiles in the $\rho-T$ diagram for the 0.89\msun \ RGB star ($\dot{M} = 1$\msun/yr, model 8) at the onset of mass transfer (solid black) and at different times during the early evolution (colors, the total mass is given in the legend). Also plotted is the location of a fixed mass coordinate (dashed black with crosses, from left to right: 0.885, 0.883 and 0.881\msun).
	\label{fig:1M_RGB_RhoT_mcoor}
	}
	\end{center}
\end{figure}

\begin{figure}[h!]
	\begin{center}
		\includegraphics[scale=0.39]{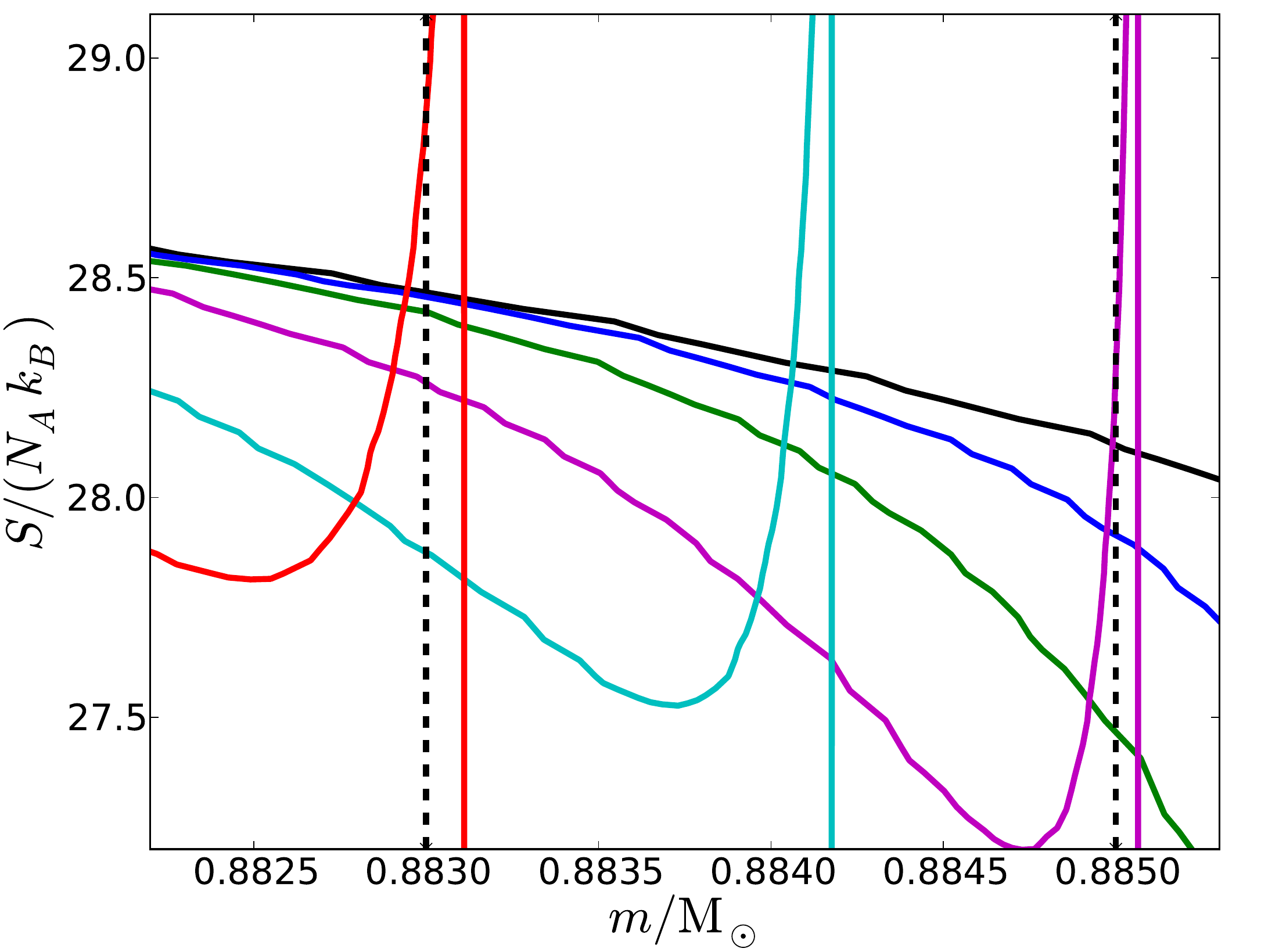}
	\caption{Entropy profiles for the 0.89\msun \ RGB star ($\dot{M} = 1$\msun/yr, model 8) at the onset of mass transfer (solid black) and at different times during the early evolution (same colors as in Figure~\ref{fig:1M_RGB_RhoT_mcoor}). Also plotted is the location of a fixed mass coordinate (dashed black, from left to right: 0.883 and 0.885\msun).
	\label{fig:1M_RGB_S_mcoor}
	}
	\end{center}
\end{figure}

We can try to explain the formation of the radiative layer mentioned above by considering a tiny ``sub-layer'' at a constant mass coordinate, located within the superadiabatic layer of the mass-losing star (Figure~\ref{fig:1M_RGB_RhoT_mcoor}). As the above layers are removed, the sub-layer can as a result more easily radiate some of its energy outwards: the temperature decreases while the local density stays almost constant. The thermal timescale is shorter in this phase than the dynamical timescale. Later on, the density profile readjusts and the density of the sub-layer drops significantly while its temperature only decreases by a small amount. Now the thermal timescale is longer than the local dynamical timescale. These two phases are also seen in the various entropy profiles (Figure~\ref{fig:1M_RGB_S_mcoor}). First, the temperature in the sub-layer decreases while the density stays almost constant, leading to a decrease of the entropy. Then, the density drops while the temperature only marginally decreases, which leads to an increasing entropy. This suggests that the readjustment of the star happens in two (nearly) distinct phases: first a thermal readjustment during which some of the energy of the layers is radiated away, then a dynamical readjustment during which the density in the outer layers decreases leading to the formation of the radiative layer.

\subsection{Additional models}

\begin{figure}[h!]
	\begin{center}
		\includegraphics[scale=0.45]{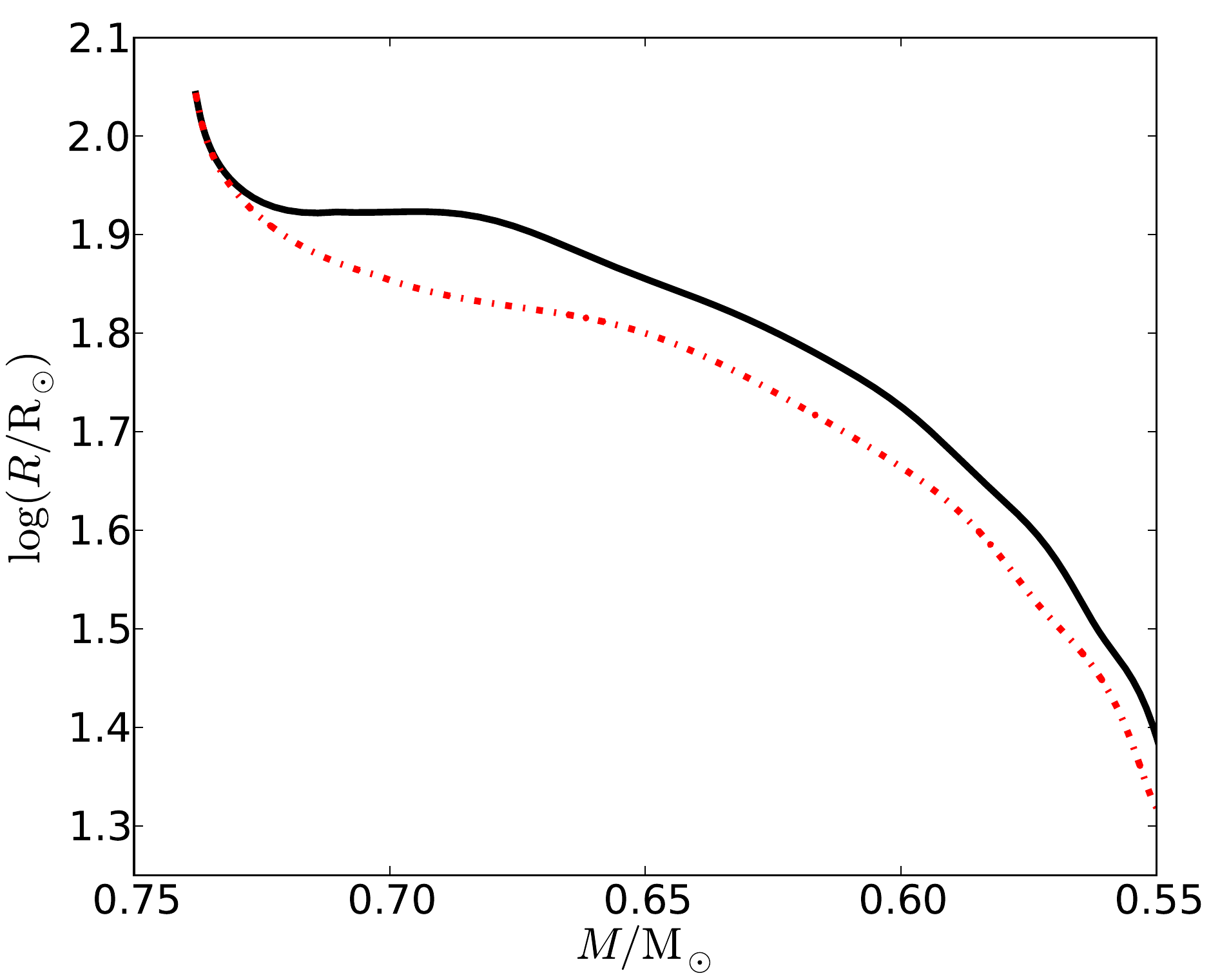}
	\caption{Evolution of the stellar radius as a function of stellar mass for the 0.74\msun \ AGB star with mass loss rates of $0.5$\msun/yr (black, model 14) and $1$\msun/yr  (dash-dotted red, model 15).
	\label{fig:1M_AGB_RadiusMass}
	}
	\end{center}
\end{figure}

\begin{figure}[h!]
	\begin{center}
			\includegraphics[scale=0.36]{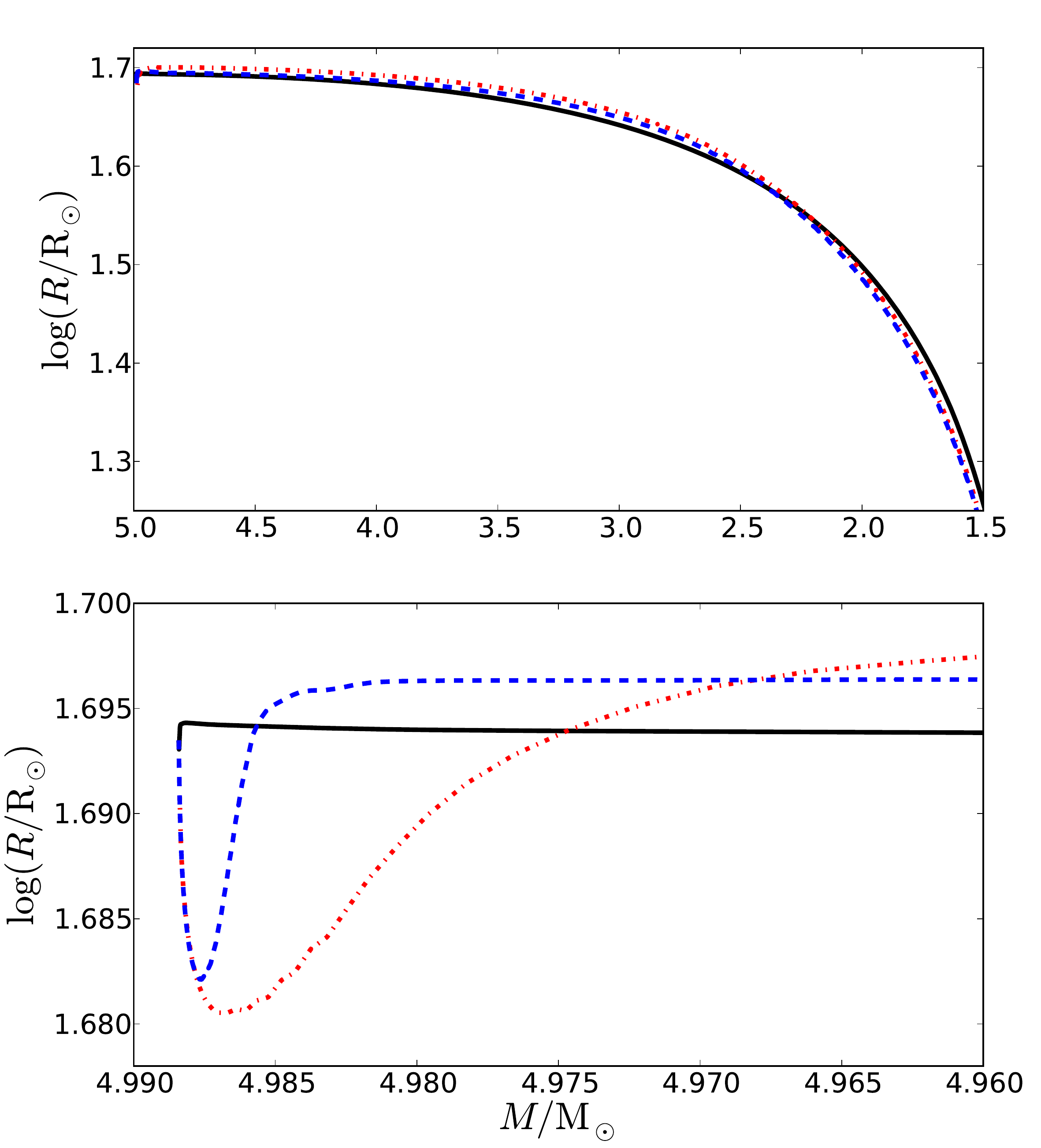}
	\caption{Top: evolution of the stellar radius as a function of stellar mass for the 5\msun \ RGB star with mass loss rates of $10^{-2}$\msun/yr (black, model 16), $1$\msun/yr  (dash-dotted red, model 17), and variable (dashed blue, model 18). Bottom: a close-up of the early evolution.
	\label{fig:5M_RGB_RadiusMass}
	}
	\end{center}
\end{figure}

We also carry out evolutionary sequences for a 0.74\msun \ AGB star with two different mass loss rates (models 14 and 15) and for a 5\msun \ RGB star (models 16, 17 and 18) similar to the one used by \cite{WoodsIvanova2011} in order to verify that the behavior seen for the 0.89\msun \ RGB star case is not a special case. The evolution of the radius for both stars is shown in Figure~\ref{fig:1M_AGB_RadiusMass} and Figure~\ref{fig:5M_RGB_RadiusMass}, respectively. Again, all the models initially shrink in radius. Models suffering a higher mass loss rate then shrink faster. In the particular case of the 5\msun \ RGB star, our results differ from the findings by \cite{WoodsIvanova2011}. Indeed, their Figure~3 shows that for all mass loss rates except the lowest one ($\dot{M} = 10^{-3}$\msun/yr), the mass-losing star slightly expands. The higher the mass loss rate, the larger the star expands. This behavior is quite similar to the one seen in Figure~\ref{fig:1M_RGB_Radius} for the hydrostatic simulations, so it is possible that the results presented in their Figure~3 have been obtained assuming hydrostatic equilibrium. Nevertheless, the stellar radius after 0.6\msun \ has been lost by the star (see their Figure~2) seems quite consistent with ours, although making a more detailed comparison is difficult as their Figure~3 only shows the evolution until the star has lost 0.5\% of its total mass. Our 5\msun \ RGB model has a smaller core mass fraction (0.122) than their model (0.171), which might also lead to differences.

\section{Summary and Discussion}
\label{sec:summary}

We used the MESA stellar evolution code to carry out one-dimensional hydrostatic and hydrodynamic simulations of mass-losing stars, either low-mass ZAMS stars between 0.3 and 0.5\msun, or 1 or 5\msun \ (main sequence mass) giant stars, with several constant and variable mass loss rates up to a few \msun/yr.

We first tested our numerical method against the low-mass ZAMS stars case and reproduced the results of \cite{GeEtAl2010}. Therefore, it is correct to assume that the evolution of the star is adiabatic {\it in this specific case}.

We then investigated the case of a 0.89\msun \ RGB star for five different mass loss rates. We showed that the mass-losing star does not remain in hydrostatic equilibrium for high mass loss rates and that the evolution is not adiabatic, as the outer superadiabatic layer has enough time to thermally relax. Only for low mass loss rates ($\dot{M} \leq 10^{-2}$\msun/yr) have both the outer superadiabatic layer and the stellar interior enough time to thermally readjust. The superadiabatic layer progressively reconstructs and survives the entire evolution, making the evolution not adiabatic both locally and globally. For high mass loss rates ($\dot{M} \geq 0.1$\msun/yr), the outer part of the star progressively loses its superadiabaticity and the interior does not have enough time to thermally readjust. Even though a fraction of the initial superadiabatic layer might survive, a larger radiative zone emerges below it and the star keeps shrinking during the entire sequence. The evolution of the star is locally non-adiabatic and hydrodynamic, as some energy that is stored in gravitational form in the hydrostatic models is actually in a kinetic form, leading to the star contracting instead of expanding.

We also carried out additional simulations for a 0.74\msun \ AGB star with a core mass of 0.52\msun\ and a 5\msun\ RGB star. These models are consistent with our previous findings and with the 5\msun\ RGB model in \cite{WoodsIvanova2011}. We have also verified that the outcomes of our simulations depend on neither numerical parameters such as the initial timestep adopted, nor on boundary conditions.

According to our stellar evolution models, giants barely expand, if at all.  This result impacts the condition for the onset of the common envelope phase. Using the Eddington luminosity limit, one can estimate the mass loss rate above which a dwarf would be unable to accrete material, to be about $10^{-3}$\msun/yr. For higher mass loss rates, the hydrostatic assumption is violated and there is no expansion of the giant's envelope. As a consequence, the positive feedback from the mass-losing giant discussed in Section~\ref{sec:intro}, may be reduced. Further investigations are required to quantify how this feedback affects the temporal evolution of the mass transfer rate. Overall, criteria for unstable mass transfer based on adiabatic mass loss models should be re-investigated. Moreover, if giant stars do not expand as a result of mass loss, this process does also not help the envelope ejection during a common envelope interaction, as speculated by \cite{AlphaPaper2011}.

\section{Acknowledgments}
\label{sec:ack}
J-CP acknowledges funding from NSF grant 0607111 and thanks Mordecai-Mark Mac Low for his support. FH acknowledges funding from an NSERC Discovery grant. BP acknowledges funding from NSF grants PHY 05-51164 and AST 07-07633. The authors are grateful to Orsola De Marco for the initial discussions regarding the stellar response to common-envelope-induced mass loss. J-CP thanks Aaron Dotter for his help installing the {\it FreeEOS} tables, and Charli Sakari for proofreading this manuscript.

\bibliographystyle{/Applications/TeX/apj}                       

\end{document}